\documentclass{article}
\usepackage{ijcai25}

\usepackage{times}
\usepackage{soul}
\usepackage{url}
\usepackage[hidelinks]{hyperref}
\usepackage[utf8]{inputenc}
\usepackage[small]{caption}
\usepackage{graphicx}
\usepackage{amsmath}
\usepackage{amsthm}
\usepackage{booktabs}
\usepackage{algorithm}
\usepackage{algorithmic}
\usepackage[switch]{lineno}
\usepackage{multirow}

\author{
Vaishali Aggarwal
\and
Sachin Thukral\and
Arnab Chatterjee\\
\affiliations
TCS Research\\
\emails
\{aggarwal.vaishali, sachi.2, arnab.chatterjee4\}@tcs.com,
}

\title{MHINDR -- a DSM5 based mental health diagnosis and recommendation framework using LLM}

\begin{document}

\maketitle

\begin{abstract}
Mental health forums offer valuable insights into psychological issues, stressors, and potential solutions. We propose MHINDR, a large language model (LLM) based framework integrated with DSM-5 criteria to analyze user-generated text, diagnose mental health conditions, and generate personalized interventions and insights for mental health practitioners. Our approach emphasizes on the extraction of temporal information for accurate diagnosis and symptom progression tracking, together with psychological features to create comprehensive mental health summaries of users. The framework delivers scalable, customizable, and data-driven therapeutic recommendations, adaptable to diverse clinical contexts, patient needs, and workplace well-being programs.
\end{abstract}

\section{Introduction}

The digital age has reshaped how people seek mental health support, with many turning to social media to share their struggles. This shift highlights the urgent need for effective tools to deliver personalized care. However, creating tailored treatment plans remains challenging due to subjective judgments and limited data. As social media becomes a key outlet for mental health discussions, efficiently extracting insights from vast data is essential for timely, appropriate interventions.

Online discussion forums are rich sources of textual data, where users share mental health concerns and receive responses, generating meaningful social engagement. This interaction adds depth to the information shared.

The DSM-5 (Diagnostic and Statistical Manual of Mental Disorders, Fifth Edition)~\cite{DSM-5} is a standardized guide used by mental health professionals. It includes detailed descriptions of various disorders and their symptoms, offering a standardized diagnostic guidance to classify mental health disorders from interviews and questionnaires.

Generative AI, particularly Large Language Models (LLMs), has emerged as a powerful tool in extracting relevant information at great details through fine tuned queries or prompts.
LLMs have shown the potential to process and extract insights from unstructured social media content, filtering out irrelevant information and focusing on mental health-related discussions. Leveraging these models, we can better understand individual users’ mental states, offering mental health practitioners valuable tools to support diagnosis and intervention. 

We present MHINDR, a Mental Health Intelligence framework for Navigating Diagnostics and Recommendation, which takes mental health related (social media) text as input, and creates a comprehensive summary of the mental health condition of an user using a reference diagnostic tool (DSM5), as well as producing recommendations for better wellbeing.

The key highlights of our framework are presented below:
\begin{itemize}
    \item  Aggregates multiple posts and comments from each individual to create a detailed mental health profile.
    \item  Utilizes the DSM-5 diagnostic criteria to classify mental health disorders and evaluate their severity, ensuring consistency with established diagnostic standards.
    \item   Extracts both temporal and non-temporal data related to the progression of mental health issues, essential for DSM-5-based diagnoses and ongoing patient evaluation.
    \item  Provides tailored recommendations to empower individuals to take proactive measures in improving their mental well-being.
\end{itemize}

This work specifically addresses user modeling, personalization and recommendation in the mental health space, for user generated textual data.
Our findings offer significant implications for mental health practitioners by providing a comprehensive end-to-end framework that offers valuable insights into users' mental health trajectories and identifying specific areas of concern that enhances treatment decisions and supports proactive mental well-being.

\section{Related work}
The analysis of social media data for mental health diagnosis has advanced significantly, leveraging platforms like Twitter and Reddit. Early efforts focused on linguistic markers of depression from Twitter~\cite{wolohan2018detecting}.
Traditional NLP methods for temporal extraction, such as TF-IDF-based logistic regression, showed strong performance in earlier studies~\cite{viani2021natural}.
The MentSum dataset~\cite{sotudeh2022mentsum} provides a valuable resource for summarizing mental health-related posts on platforms like Reddit, enabling concise yet informative summaries that are useful for mental health counselors. The ``Dreaddit” dataset, designed for stress analysis in social media posts, demonstrated the efficacy of NLP techniques in identifying stress-related content on Reddit through binary classification~\cite{turcan2019dreaddit}. Furthermore, Ref.~\cite{sampath2022data} developed a gold standard dataset for detecting depression levels from social media posts, employing traditional algorithms and data augmentation within a multi-class classification framework.
Machine learning methods, including deep learning, have improved detection accuracy on social media, identifying disorders such as depression, anxiety, and bipolar disorder from Reddit posts~\cite{kim2020deep}. Efforts to classify posts into DSM-5 categories exist~\cite{gaur2018let}, but they often neglect temporal dynamics, limiting their clinical applicability.

Recent advancements~\cite{yang2024mentallama} involve large language models (LLMs), which excel in context-sensitive tasks. For instance, LLMs combined with hierarchical VAEs can summarize mental health timelines~\cite{song2024combining}, while prompt engineering enhances their ability to address diverse tasks~\cite{wang2023element}. 
Additionally, event extraction have been cast as a machine reading comprehension (MRC) problem  by constructing questions to query event triggers and arguments~\cite{liu2020event,du2020event,li2020event}. A range of machine learning techniques has been employed to detect mental health issues on social media platforms, with deep learning methods significantly enhancing detection accuracy~\cite{tadesse2019detection}. Tools like “Illuminate” leverage advanced LLMs for diagnosis and therapy recommendations~\cite{agarwal2024illuminate}. Computational frameworks further evaluate LLMs as potential therapists~\cite{chiu2024computational}.

\section{Dataset}
Reddit, a popular social discussion platform for diverse and active communities, organizes content into subreddits based on various domains. For this study, we extracted text data from users' posts and comments, along with metadata indicating whether the entry was a post or a comment, as well as the creation date and time. The dataset was sourced from subreddits \texttt{r/mentalhealth} and \texttt{r/depression} over a one-year period, spanning between December 1, 2019 and November 30, 2020, a total of $334,197$ posts and $1,063,507$ comments. This dataset was gathered using the Pushshift~\cite{Reddit_dump} API, providing a comprehensive foundation for our analysis.

\section{Methodology}
\label{sec:methodology}
\begin{figure*}[t]
    \centering
    \includegraphics[width =0.9\linewidth]
    {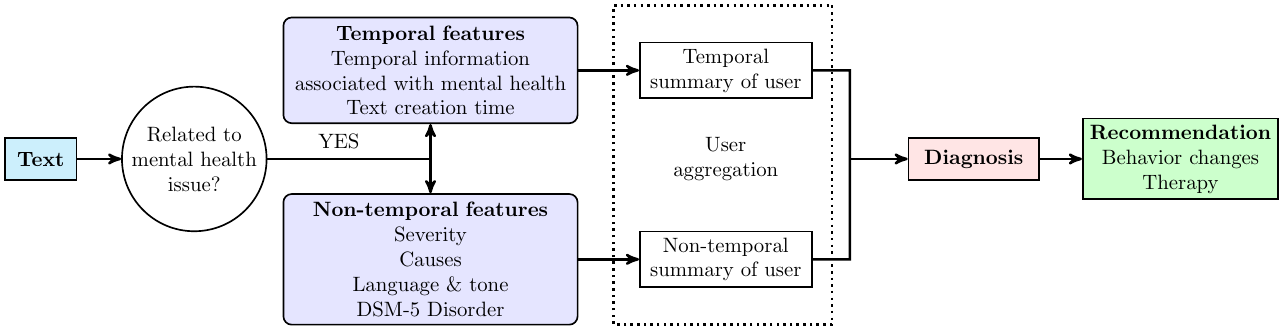}
    \caption{MHINDR -- an end-to-end framework for generating user summary and recommendation from text input data}
    \label{fig:pipeline}
\end{figure*}

The MHINDR framework shown in Fig.\ref{fig:pipeline} details the process from data filtration to user aggregation, leading to personalized mental health summaries and recommendations. We use Llama3.1\cite{dubey2024llama} via the Groq API\footnote{https://groq.com/groqcloud/} (LLM) for all tasks unless specified, achieving sub-second processing times. The LLM is configured with temperature=0, max\_tokens=1000, top\_p=1.0, and no stop sequence to ensure deterministic, comprehensive outputs.

\subsection{Data Filtering}
\label{subsec:preprocessing}
The first step in the process is data pre-processing to ensure a clean, structured dataset. This includes removing irrelevant content (e.g., deleted posts, image links), noise (e.g., special characters, punctuation, URLs, HTML tags), and correcting formatting issues for consistency and clarity.

An additional filtering process uses LLM to classify each post with a strict binary “yes” or “no” to filter content genuinely relevant to mental health, ensuring the final dataset's integrity and focus.

\subsection{Feature Extraction}
\label{subsecm:symptom}
In our approach, we extracted two broad types of features: non-temporal capture the basic aspects of the mental health of the user and temporal temporal such as dates and durations, crucial for capturing event sequences and behavioral patterns. 
The temporal information present in the dataset was notably sparse, posing significant challenges in capturing it directly from the text. To address this issue, we processed both types separately using tailored prompts, generating two distinct summaries per user.

\subsubsection{Non-Temporal Feature Extraction}
For each entry, the following non-temporal features were extracted:
\begin{enumerate}
    \item \textbf{Severity Assessment}: Classifying disorders as \textit{mild}, \textit{moderate}, or \textit{severe} to understand issue intensity using DSM-5~\cite{DSM-5} diagnostic tool.
    \item \textbf{Identification of Causal Factors}: Identifying stressors or life events impacting mental health.
    \item \textbf{Language and Tone Analysis}: Evaluating text for emotional intensity, mood changes, and well-being.
    \item \textbf{DSM-5-Based Disorder Classification}: Classifying disorders using DSM-5~\cite{DSM-5} criteria for a systematic approach.
\end{enumerate}

\subsubsection{Temporal Feature Extraction}
Temporal features can improve the effectiveness of the DSM-5 psychiatric tool in providing diagnosis and treatment recommendations. 
For each entry , we have two timeline information:
\begin{enumerate}
    \item \textbf{Text Creation Time}: The timestamp indicating when the post was created, as recorded in the dataset.
    \item \textbf{Temporal Information}: We extracted presence of any temporal references within the user's text using prompts. In cases where no such information was found, the entry was assigned a value of "No Timeline".
\end{enumerate}
This method facilitated a more structured analysis of the temporal aspects within the data.

\subsection{User Aggregation and Mental Health Summary}
\label{subsecm:aggregation}
After extracting temporal and non-temporal features, we aggregate each user's posts and comments into a comprehensive feature set. 
This aggregation includes both temporal and non-temporal features, allowing for an in-depth analysis of the user’s profile. 
This enables detailed user profiling to support personalized mental health summaries. 

Our framework separately extracts temporal and non-temporal summaries using tailored prompts to reduce bias and improve accuracy.  By separating the summaries, we ensure that each summary is tailored to its specific context, resulting in a more reliable and informative overview of the user’s mental health status. Ultimately, integrating the insights from both summaries will yield a concise and overall summary for the user.

\subsubsection{Non-Temporal Summary}
The non-temporal summary is derived from aggregated user posts and focuses on key aspects:
\begin{enumerate}
    \item[(i)] overall disorder severity,
    \item [(ii)] major mental health triggers,
    \item[(iii)] prominent DSM-5 disorders,
    \item[(iv)] predominant language style and tone,
    \item[(v)] recurring themes observed in the user's texts,
    \item[(vi)] overall mental health status of the user.
\end{enumerate}

\subsubsection{Temporal Summary}
Temporal summary extraction was challenging due to confusion between post dates and in-text timelines, and frequent "No Timeline" labels. To resolve this, we aggregated posts with temporal cues per user, creating a chronological sequence based on dates, extracted timelines, and content.
This sequence reveals temporal patterns for each user, capturing:
\begin{enumerate}
    \item[(i)] Duration of issues,
    \item[(ii)] Frequency of posting or mentions of issues,
    \item[(iii)] Recurrence of concerns,
    \item[(iv)] explicit mentions of time periods.
\end{enumerate}

\subsection{Diagnosis}
In the next phase, we combined both temporal and non-temporal summary features to create a comprehensive mental health profile for each user. This integrated summary consolidates all relevant features, highlights critical areas of concern, and offers a structured view of the user’s mental health journey. It provides valuable insights into the user’s emotional and psychological well-being over time, serving as a foundation for generating precise, actionable mental health recommendations.

\begin{table*}[h]
\caption{Prompts for final diagnosis and providing recommendation using the diagnostic dataframe as an input for the user}
\label{table:prompts3}
\begin{tabular}{|l|p{7.5cm}|p{6.3cm}|}
\hline
Task &
  System &
  User \\ \hline
  Diagnosis &
  You are an advanced language model trained to analyze and summarize text data related to mental state of a particular user. Given a data covering various features of a user based on the content posted by him/her on reddit, your task is to further summarize it concisely and provide the summary limited to 400 words covering every aspect with clear information such that when the summary is given to a mental health practitioner who works in accordance with DSM-5, it becomes helpful for him/her to recommend the therapy and behavior changes to the user based on it. Your task is specified to generate such a summary which covers all the necessary information regarding the mental health state of the user. &
  Summary : [Dataframe] Review the provided data of the user's overall mental state covering several features. Summarize this further concisely and provide the summary in 400 words covering every aspect with clear information such that when the summary is given to a mental health practitioner who works in accordance with DSM-5, it becomes helpful for him/her to recommend the therapy and behavior changes to the user based on it. Your task is specified to generate such a summary which covers all the necessary information regarding the mental health state of the user. \\ \hline
Recommendations &
  You are a Mental Health Practitioner working in accordance with DSM-5. Given the summary of a user's behavior and mental health trends, provide therapy recommendations and suggest behavior changes separately that could benefit the user. Consider all the relevant and important information mentioned in the summary and recommend at the max three most suitable therapies and five actionable behavior changes accordingly. &
  Summary : [Dataframe] Review the provided summary of the user's behavior and mental health. Consider all the important information, and generate at the max three most suitable therapy recommendations and five actionable behavior modification suggestions separately that could help the user improve their well-being. Do not provide any extra explainations. \\ \hline
\end{tabular}
\end{table*}

\subsection{Recommendation}
\label{subsecm:recom}

\textbf{Suitable Therapy}: 
This module is designed to offer personalized therapy recommendations harnessing the capabilities of LLM. Utilizing the mental health summary generated for each user in the previous step, the LLM were tasked to identify the most appropriate therapeutic interventions for him/her. The objective is to provide users with insights into their mental state and recommend therapies that are most likely to be effective for them. By integrating this module, we aim to deliver personalized, data-driven therapeutic options, thus improving the effectiveness and precision of mental health support.

\noindent \textbf{Behaviour Change}: 
The recommendation module leverages LLM to provide actionable behavior change suggestions that improve mental well-being. Based on the user's summary, LLM offers practical and tailored advice to support mental health improvement, complementing therapeutic recommendations for a holistic approach to care.

The prompts for the diagnosis and recommendations tasks are provided in Table~\ref{table:prompts3}.

\section{Results}
We curated a dataset of 36,890 posts and comments from 200 active users, selected by posting frequency. After filtering, 15,474 entries were retained for analysis. Results for the features outlined in Sec.\ref{sec:methodology} are presented in the following subsections, with examples shown in Tables~\ref{table:examples} and~\ref{table:examples1}.

\subsection{Severity}

Table~\ref{table:severity} highlights that $67.4\%$ of entries fall under the \textit{Severe} category, $19.4\%$ are \textit{Moderate}, and only $1.9\%$ are \textit{Mild}, with $11.3\%$ uncategorized due to ethical concerns over mentions of suicide or self-harm.

At the user level, 58.3\% are in the *moderate-to-severe* range, 40.3\% in \textit{severe}, and 1\% in \textit{mild-to-moderate}. The high proportion of \textit{severe} cases highlights the need for specialized care, while the prevalence of \textit{moderate-to-severe} users emphasizes the importance of timely interventions. The low percentage of \textit{mild-to-moderate} cases points to potential gaps in early detection and the need for improved screening and awareness.

\begin{table}[h]
\caption{Analysis of Severity Level Classification}
\label{table:severity}
\resizebox{\linewidth}{!}{%
\centering{
\begin{tabular}{|l|l|l|l|l|}
\hline 
\multirow{2}{*}{Entries} & Mild & Moderate & Severe & Extreme\\\cline{2-5} 
                  & $1.9\%$ & $19.4\%$ & $67.4\%$ & $11.3\%$ \\ 
\hline  \hline 
\multirow{2}{*}{Users} & Mild-to-Moderate & Moderate-to-Severe & Severe & \\\cline{2-5} 
                   & $1\%$ & $58.3\%$ & $40.7\%$ & \\ \hline
\end{tabular}
}
}
\end{table}

\subsection{Causes}
Our framework identifies key causes of mental health issues, including social isolation, low self-esteem, negative self-talk, childhood trauma, emotional distress, hopelessness, lack of support, sleep disturbances, and financial stress, which frequently appear across user posts.

Identifying causes of mental health issues is crucial for addressing root problems rather than just symptoms. For each user, we compile a list of mental health triggers from their posts. For example, Table~\ref{table:examples} shows financial insecurity and unemployment as key triggers for one user, helping practitioners design targeted, effective interventions.

\subsection{Language Style and Tone}
Analysis of user posts revealed prevalent emotional tones such as \textit{despondent}, \textit{melancholic}, and \textit{desperate}, reflecting emotional lows and helplessness. Tones like \textit{vulnerable}, \textit{frustrated}, and \textit{introspective} highlight struggles with self-criticism. The frequent \textit{confessional} tone indicates social media's therapeutic role, while tones like \textit{resigned}, \textit{isolated}, and \textit{sad} emphasize loneliness and mental health decline. Negative tones strongly correlate with depressive disorders, though occasional positive or neutral tones, such as \textit{empathetic}, \textit{sarcastic} and \textit{emotive}, suggest attempts to connect or express emotions.

\begin{table*}[t]
\caption{Extracted temporal and non-temporal summary for a sample user from all its posts and comments}
\label{table:examples}
\begin{tabular}{|p{2.0cm}|p{1.9cm}|p{12.4cm}|}
\hline
 &
  Features &
  LLM Output \\ \hline
\multirow{6}{*}{\begin{tabular}[c]{@{}l@{}}Non-Temporal\\ Features \\Summary\end{tabular}} &
  Severity &
  Moderate to Severe \\ \cline{2-3} 
 &
  Causes &
  Financial insecurity, poor eating habits, social isolation, lack of agency, job dissatisfaction, unemployment, corporate stress, lack of social life, lack of skills, low self-esteem, negative self-talk, pessimism, uncertainty, online presence, perceived threats, lack of emotional regulation skills, childhood trauma, parental influence \\ \cline{2-3} 
 &
  Language Style and Tone &
  The user's language style is confessional, introspective, and self-deprecating, with a tone that is often melancholic, anxious, and frustrated. \\ \cline{2-3} 
 &
  Recurring Themes &
  The user's content frequently references themes of self-doubt, low self-esteem, and feelings of inadequacy, as well as struggles with social anxiety, depression, and unemployment. \\ \cline{2-3} 
 &
  Disorders &
  The user's symptoms align with the DSM-5 criteria for Major Depressive Disorder, Generalized Anxiety Disorder, Social Anxiety Disorder, Avoidant Personality Disorder, Borderline Personality Disorder, Schizotypal Personality Disorder, and Schizophrenia. \\ \cline{2-3} 
 &
  Mental Health Status &
  The user's mental health condition is characterized by a chronic and worsening trajectory, with periods of remission and relapse. The condition has a significant impact on the user's daily life, relationships, and overall well-being. \\ \hline
\multirow{2}{*}{\begin{tabular}[c]{@{}l@{}}Temporal \\ Features\\ Summary\end{tabular}} &
  Chronological Events &
  December 14, 2019: The user reports having a "shitty personality" and poor mental health, with a timeline of "past half decade" indicating that the issue has been ongoing for at least 5 years. December 28, 2019: The user reiterates their poor mental health and "shitty personality", with a timeline of "5 years" further emphasizing the duration of the issue. \\ \cline{2-3} 
 &
  Temporal Patterns &
  The temporal patterns observed in the dataset are: Duration: The user reports a duration of at least 5 years for their poor mental health and "shitty personality". Frequency: The user reports on their mental health issues twice, with a gap of 14 days between the two posts. Recurrence: The user mentions a recurring pattern of poor mental health and "shitty personality" over a prolonged period. Specific mentions of time: The user mentions "past half decade" and "5 years" in the text, indicating a specific timeframe for their mental health issues. \\ \hline

\end{tabular}
\end{table*}

\begin{table*}[t]
\caption{Example diagnosis and recommendations for a sample user}
\label{table:examples1}
\begin{tabular}{|p{2.5cm}|p{1.4cm}|p{12.5cm}|}
\hline
 &
  Features &
  LLM Output \\ \hline
  DSM-5 diagnosis &
   &
  Based on the provided data, user's mental state is characterized as moderate to severe, with symptoms aligning with the DSM-5 criteria for a mental health disorder. The user's language style is confessional, indicating a willingness to share personal struggles and emotions. Recurring themes in their content include references to financial insecurity, poor eating habits, and social issues, which may be contributing factors to their mental health condition. The user's mental health status is characterized as characterized by symptoms of anxiety, depression, and possibly trauma. The temporal patterns observed in their posts suggest a worsening of symptoms over time, with a notable increase in distressing language and emotional expression in the second post. This may indicate a growing sense of hopelessness and despair. The user's mental health condition is likely to be complex and multifaceted, with potential underlying causes including financial insecurity, poor eating habits, and social issues. The confessional language style and recurring themes in their content suggest a deep-seated sense of vulnerability and emotional pain. A mental health practitioner working in accordance with DSM-5 criteria may consider the following potential diagnoses: Major Depressive Disorder, Generalized Anxiety Disorder, or Post-Traumatic Stress Disorder. \\ \hline
\multirow{2}{*}{Recommendations} &
  Therapy &
  1. Cognitive-Behavioral Therapy (CBT) 2. Interpersonal Therapy (IPT) 3. Trauma-Focused Cognitive-Behavioral Therapy (TF-CBT) \\ \cline{2-3}  
 &
  Behavior Change &
  1. Establish a consistent sleep schedule and prioritize self-care activities, such as meditation, yoga, or reading, to improve emotional regulation and reduce symptoms of anxiety and depression. 2. Create a budget and prioritize financial planning to reduce financial insecurity and stress. 3. Engage in regular physical activity, such as walking or jogging, to improve mood and reduce symptoms of depression. 4. Practice mindful eating habits, such as meal planning and preparation, to improve nutrition and reduce symptoms of depression. 5. Set boundaries and prioritize social connections with supportive individuals to reduce feelings of loneliness and isolation. \\ \hline
\end{tabular}%
\end{table*}

\subsection{DSM-5 Disorders}
The analysis identifies a range of DSM-5 disorders, with Major Depressive Disorder being the most common, followed by Borderline Personality Disorder and Suicidal Ideation. Anxiety-related disorders like Social Anxiety and Generalized Anxiety Disorder also appear frequently. Other conditions such as Adjustment Disorder, PTSD, and OCD highlight struggles with trauma and compulsive behaviors, while less common disorders like Depersonalization, Existential Crises, and Body Dysmorphic Disorder reflect identity and appearance anxieties. These findings emphasize the need for a comprehensive support framework addressing both prevalent and less common mental health challenges for more personalized interventions.

The identification of high-frequency disorders highlights the need for a balanced approach to mental health support. Practitioners and researchers must address both common and less frequent, yet impactful, mental health challenges. This analysis offers valuable insights into the spectrum of issues discussed in online forums, aiding in the development of more personalized and effective interventions.

\subsection{Temporal Information}
Only 10.65\% of the posts contained explicit temporal references, limiting the generation of temporal summaries. This highlights the challenge of extracting time-related context from user content. Despite this, posts with temporal data provided valuable insights into the duration and recurrence of mental health issues. Temporal summaries were generated for 92.46\% of users, demonstrating that most shared relevant time-related details. Table~\ref{table:examples} illustrates how scattered temporal clues can be combined into a cohesive mental health timeline.

\subsection{Diagnosis}
As shown in Table~\ref{table:examples1}, the final diagnosis integrates both \textit{temporal} and \textit{non-temporal} information, providing a comprehensive overview essential for mental health practitioners.
Temporal data captures the progression of symptoms, highlighting patterns such as fluctuations, stability, or worsening, and the duration of issues. This dynamic insight helps identify critical periods for timely intervention. Non-temporal data includes static factors like demographics, medical history, and environmental influences, offering context to interpret temporal changes and personalize treatment.

This dual-layered approach enables practitioners to address current symptoms while anticipating future challenges, creating a personalized and proactive care plan tailored to the user's needs and long-term well-being.

\subsection{Recommendations}

We focused on two key areas -- therapy recommendations and behavior change suggestions, both derived from the DSM-5 diagnosis summary generated by the LLM. For therapy recommendations, the LLM was asked to recommend the three most suitable therapies for the users based on their diagnosed mental state. 

It is observed that \textit{Cognitive Behavioral Therapy (CBT)} emerged as the most common recommendation, suggested for $185$ out of $200$ users, highlighting its versatility in addressing various mental health issues. \textit{Dialectical Behavior Therapy (DBT)} followed with $118$ recommendations, emphasizing its role in emotional regulation and interpersonal skills, while \textit{Interpersonal Therapy (IPT)} was suggested for $112$ users, focusing on improving social functioning and relationships. Beyond these, a variety of therapies were tailored to meet users' specific needs, showcasing the importance of personalized treatment in mental health care.

Additionally, the LLM provided practical behavior change suggestions, including \textit{mindfulness practices}, establishing healthy routines, and coping strategies. These actionable recommendations promote a positive mindset and empower users to enhance their well-being. For instance, Table~\ref{table:examples1} highlights advice such as maintaining a regular sleep schedule and engaging in self-care activities like meditation, yoga, or reading. These strategies address immediate emotional needs while fostering long-term mental health and personal growth.

\section{User Interaction Relation}
We also analyzed Reddit post-comment interactions to assess relationship types using an LLM. Each pair was evaluated for relevance and categorized, with unrelated pairs labeled “Not Related.” Overall, 87.27\% of pairs were related, indicating high engagement. Due to ethical restrictions, 6.9\% of posts involving self-harm or suicidal thoughts could not be processed.

Empathy emerged as the most common relationship type, followed by agreement, support, shared experiences, criticism, and encouragement. These findings reveal how users engage emotionally, offering valuable insights into communication patterns relevant to mental health and social behavior.

\section{Ethical Concerns and Scalability}
Ethical considerations guide the process, with LLMs handling sensitive content carefully and avoiding outputs related to suicide or self-harm to ensure user safety. Additionally, our framework is designed to be inherently scalable, ensuring that, given sufficient computational resources, it can efficiently handle large-scale datasets. This scalability ensures that our framework remains practical for real-time applications in psychiatric assessment and mental health screening, supporting widespread adoption and integration with existing clinical systems.

\section{Conclusion}
This paper introduces MHINDR, an end-to-end framework for detecting and diagnosing mental health disorders using DSM-5 criteria and generating personalized recommendations. While we used DSM-5 as the primary diagnostic tool, the framework can be adapted to other psychiatric guidelines as well.
We analyzed Reddit posts and comments, extracting both temporal and non-temporal features to generate separate summaries. Non-temporal cues (e.g., severity, language, potential causes) enriched diagnosis, while temporal data captured symptom progression over time, offering deeper insight into users’ mental health.

LLMs tuned for complex tasks, enabled efficient analysis of large-scale social discussion data while preserving focus on individual user experiences, significantly streamlined our workflow, accelerating tasks like feature extraction and aggregation compared to traditional multi-model approaches. 

Our framework combines the outputs (see e.g., Table~\ref{table:examples}) to create a nuanced profile of each user, allowing for diagnosis and tailored recommendations (see e.g., Table~\ref{table:examples1}) that address both their current state and future risks.
This approach supports mental health practitioners by informing targeted interventions while allowing flexibility for expert adjustments. Results highlight the potential of leveraging public social media data to deliver scalable, personalized mental health care with actionable insights.

For us, validation remains a crucial aspect of future work, to ensure the framework’s effectiveness, reliability, and alignment with established psychiatric practices before widespread adoption in clinical settings.

Our work lays the foundation for future research in utilizing online forums for mental health diagnostics and recommendation systems, opening up possibilities for early detection and timely intervention in mental health care. This framework, developed on static conversational data, can be naturally adapted to real-time textual interaction between user and mental health practitioner or even user and AI-powered interface (chatbot) in the mental health domain.


\bibliographystyle{ijcai25}
\bibliography{ijcai25}

\end{document}